\documentclass[twocolumn,twoside,slac_two]{revtex4}
\usepackage{graphicx}
\usepackage{fancyhdr}
\usepackage{hyperref}
\hypersetup{
    colorlinks=true,
    urlcolor=magenta,
}
\begin{document}

%Title of paper
\title{The Tunka Radio Extension, an antenna array for high-energy cosmic-ray detection}

\author{Y. Kazarina\textsuperscript{a}, P.A. Bezyazeekov\textsuperscript{a}, N.M. Budnev\textsuperscript{a}, O. Fedorov\textsuperscript{a}, 
O.A. Gress\textsuperscript{a}, A. Haungs\textsuperscript{b}, R.~Hiller\textsuperscript{b}, T. Huege\textsuperscript{b}, 
M. Kleifges\textsuperscript{c}, E.E. Korosteleva\textsuperscript{d},
D. Kostunin\textsuperscript{b}, O. Kr\"omer\textsuperscript{c}, V. Kungel\textsuperscript{b}, L.A.~Kuzmichev\textsuperscript{d}, N. Lubsandorzhiev\textsuperscript{d}, T. N. Marshalkina\textsuperscript{a}, R.R. Mirgazov\textsuperscript{a}, R. Monkhoev\textsuperscript{a}, E.A. Osipova\textsuperscript{d},
A. Pakhorukov\textsuperscript{a}, L. Pankov\textsuperscript{a}, V.V. Prosin\textsuperscript{d}, F.G. Schr\"oder\textsuperscript{b}, A. Zagorodnikov\textsuperscript{a}
}

\affiliation{
\textsuperscript{a}Institute of Applied Physics ISU, Irkutsk, Russia;
}
\affiliation{
\textsuperscript{b}Institut f\"ur Kernphysik, Karlsruhe Institute of Technology (KIT), Germany;
}
\affiliation{
\textsuperscript{c}Institut f\"ur Prozessdatenverarbeitung und Elektronik, KIT, Germany;
}
\affiliation{
\textsuperscript{d}Skobeltsyn Institute of Nuclear Physics MSU, Moscow, Russia;
}

\begin{abstract}
The Tunka-Rex (Tunka Radio Extension) has been deployed in autumn 2012 at the territory 
of the Tunka-133 experiment (Tunka Valley, Republic of Buryatia, Russia), covering an 
area of approximately 3 km$^2$. Tunka-133 detects the Cherenkov radiation from air showers 
of cosmic rays at energies 10$^{16.5}$−--10$^{18}$ eV, and 63 antennas of Tunka-Rex measure the 
radio emission of the same air showers. Three years of joint operation of Tunka-Rex and Tunka-133 
have shown that a calibrated radio array can be used for an independent test of the scale of 
the cosmic-ray energy. Furthermore, by direct comparison of the depth of the shower maximum 
measured by Tunka-133 and Tunka-Rex, it was shown that the precision of the radio technique 
for the shower maximum is at least 40 g/cm$^2$. Two thirds of antennas are connected to the 
recently deployed arrays of scintillation stations Tunka-Grande. As next step the cross-calibration 
of Tunka-Rex and Tunka-Grande is planned, which provides the possibility of the combined measurements 
of the muon and electromagnetic components of air-showers, where the radio array will provide sensitivity 
to the shower maximum with full duty-cycle. Exploiting the complementary muon/radio information, it should 
be possible to improve the mass separation in cosmic-ray spectra. This article presents the first results 
of the combined measurements of Tunka-Rex and Tunka-Grande as well as studies of the antenna alignment 
effect and an overview of the recent Tunka-Rex results.
\end{abstract}

\maketitle

\thispagestyle{fancy}

\section{INTRODUCTION}
\begin{figure}
\includegraphics[width=1.0\linewidth]{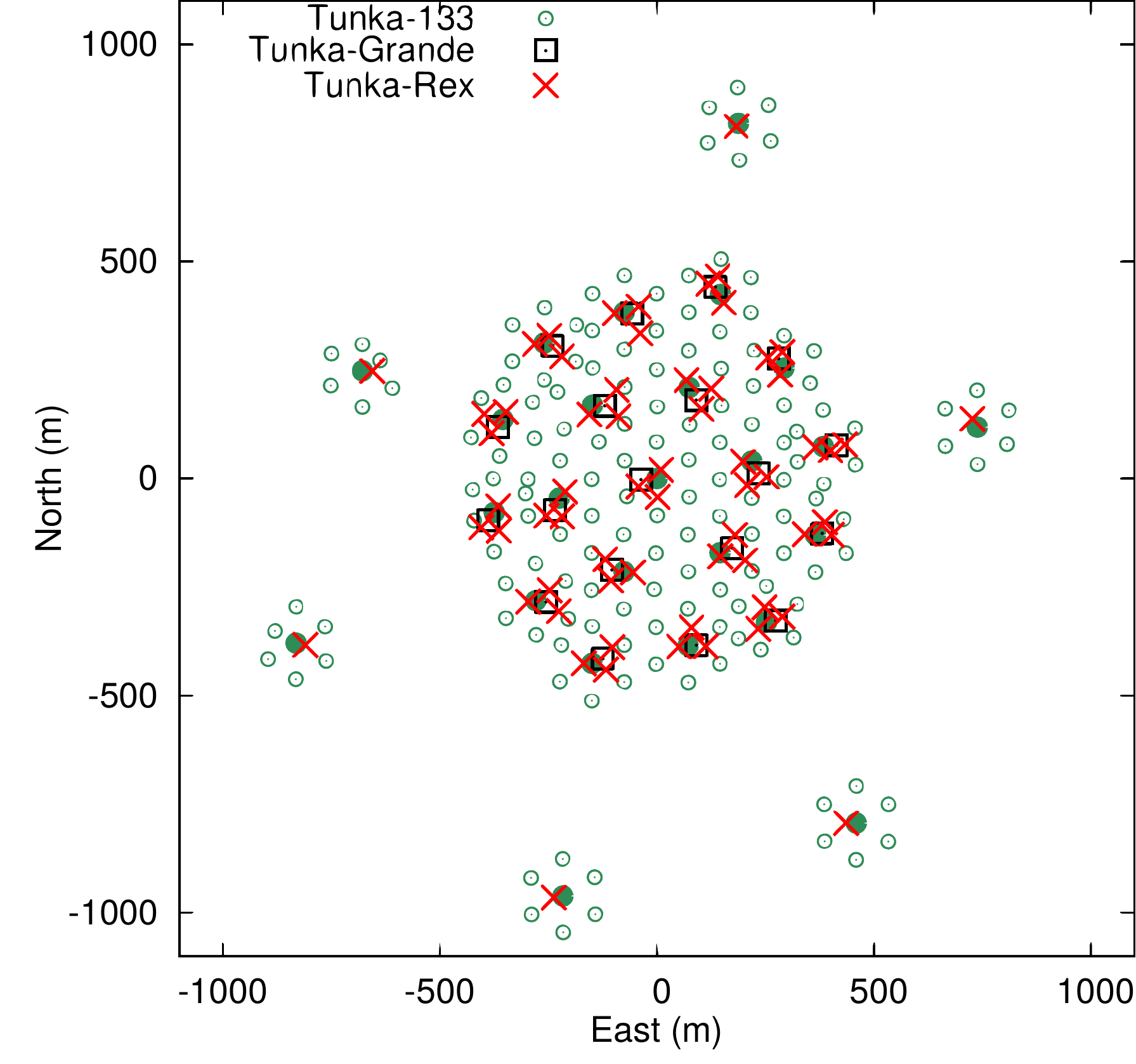} 
\caption{Layout of cosmic ray experiments of TAIGA observatory.}
\label{fig:map}
\end{figure}

One of the main puzzles of modern astrophysics are the sources of cosmic rays and their acceleration mechanisms. 
The study of cosmic rays in the energy range from 10$^{16}$ to 10$^{19}$ eV is of special interest. 
In this range a transition from galactic to extragalactic sources is supposed~\cite{Apel2012183, Sveshnikova:2015vga, SchroederReview2016}. 
Good sensitivity to the mass composition of the primary cosmic rays is required for the study of this part of the spectrum.  
There are two principle methods for cosmic ray detection: direct (satellite detectors measuring 
primary cosmic rays while orbiting Earth) and indirect (ground arrays measuring extensive air-showers 
(EAS) produced by high-energy cosmic rays) for energies above 10$^{14}$ eV, since the flux of the primary 
cosmic rays becomes too low for direct measurements. 
Indirect methods for the study of cosmic rays are used at the Tunka Valley at the observatory TAIGA 
(Tunka Advanced Instrument for cosmic ray physics and Gamma Astronomy)~\cite{TAIGA_2014}. It is a complex, hybrid 
detector, based on arrays of different types. 
There are three detectors for cosmic rays: the air-Cherenkov detector Tunka-133, the radio detector 
Tunka-Rex and the particle detector Tunka-Grande. 
TAIGA includes low threshold gamma ray detectors as well: the non-imaging air-Cherenkov detector 
Tunka-HiSCORE and Imaging Atmospheric Cherenkov Telescopes. 
The three experiments, namely Tunka-133~\cite{Antokhonov:2011zza}, Tunka-Grande~\cite{Budnev:2015xca} and Tunka-Rex~\cite{TunkaRex_NIM_2015}, conduct joint 
measurements of showers from primary cosmic rays with energies from 10$^{16}$ to 10$^{18}$ eV. FIG.~\ref{fig:map}
shows the layout of TAIGA cosmic ray detectors, which are distributed over 3 km$^2$.

\section{TUNKA-REX}
The Tunka Radio Extension (Tunka-Rex) is an array of 63 antennas distributed on an area of 3 km$^2$. 
Of them 57 antennas are occupying a denser part of the detector, an area with radius of 500 m, 
and 6 satellite antenna stations are placed at a distance of 1 km from the center of the setup. 
The central antenna stations are grouped in 19 clusters of 3 antennas each with distances between 
the cluster centers of 200 m. The layout of the setup is given in FIG.~\ref{fig:map}. Each Tunka-Rex antenna 
station consists of two perpendicular short aperiodic loaded loop antennas (SALLA)~\cite{KroemerSALLAIcrc2009, Abreu:2012pi}. Before digitalization, signals are analogically 
pre-amplified by a low noise amplifier and processed with a filter-amplifier with an effective 
band of 30-76 MHz. Each Tunka-Rex antenna station is connected either to the Tunka-133 or the 
Tunka-Grande local data acquisition and shares the same ADC boards. The frequency band of Tunka-Rex 
provides a high signal-to-noise ratio (SNR), and the atmosphere is transparent for these radio frequencies. 

%\begin{figure}
%\includegraphics[width=1.0\linewidth]{antenna.jpg}
%\caption{Tunka-Rex antenna station.}
%\label{fig:antenna}
%\end{figure}

\section{THE MAIN TUNKA-REX RESULTS}
The Tunka-Rex reconstruction methods were developed and applied for the first two seasons of 
Tunka-Rex and Tunka-133 joint measurements~\cite{Bezyazeekov:2015ica}. Only events with energies above 10$^{17}$ eV are 
taken for the analysis. The reconstruction of shower parameters is based on the lateral distribution, 
i.e., the dependence of the radio amplitude on the distance to the shower axis. The amplitude parameter 
of the lateral distribution is correlated with the primary energy and the slope of the lateral 
distribution is sensitive to the position of the shower maximum, $X_{max}$,~\cite{Kostunin:2015taa}. The energy and $X_{max}$
reconstructed by Tunka-Rex have a strong correlation with the same parameters reconstructed by 
Tunka-133. The achieved resolution for the energy reconstruction for Tunka-Rex is 15\%. 
When exploiting the shower geometry reconstructed by the host detector Tunka-133, 
the energy can be estimated even with a single antenna station to about 20\% 
precision~\cite{HillerArena2016}. The $X_{max}$ resolution of Tunka-Rex is approximately 40 g/cm$^2$ - 
for high-quality events. This can be improved by increasing the number of 
antennas and by a stricter event selection using high quality cuts. It means that 
the shower reconstruction by Tunka-Rex is reliable and provides a similar precision 
as other modern radio experiments (AERA~\cite{Aab:2015vta}, LOFAR~\cite{NellesLOFAR_calibration2015, Buitink:2014eqa}) and 
as the established air-shower techniques.
One of the main Tunka-Rex results is the energy scale comparison of the KASCADE-Grande~\cite{Apel:2010zz} and 
Tunka-133~\cite{Sveshnikova:2015vga} experiments via their radio extensions, LOPES~\cite{2014ApelLOPES_MassComposition} 
and Tunka-Rex, respectively~\cite{TunkaRexScale2016}. We used two different analysis approaches: one relying purely on measurements, 
the other one using CoREAS simulations for comparison. With both approaches it was consistently 
shown that the energy scale of the cosmic rays measurements in KASCADE-Grande and Tunka-133 
experiments agree with each other within a relative uncertainty of about 10\%.
\section{FIRST RESULTS OF JOINT MEASUREMENTS OF TUNKA-REX AND TUNKA-GRANDE}
\begin{figure}
\includegraphics[width=1.0\linewidth]{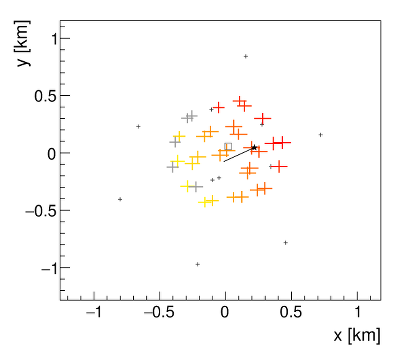} 
\caption{Example of a reconstructed event: footprint of the event, where the size of the crosses indicates the signal strength in each polarization, the color code is the arrival time, and the line and star are the direction and shower core, respectively. Grey stations are below threshold, and small crosses indicate stations not operating during this event.}
\label{fig:event}
\end{figure}

\begin{figure}
\includegraphics[width=1.0\linewidth]{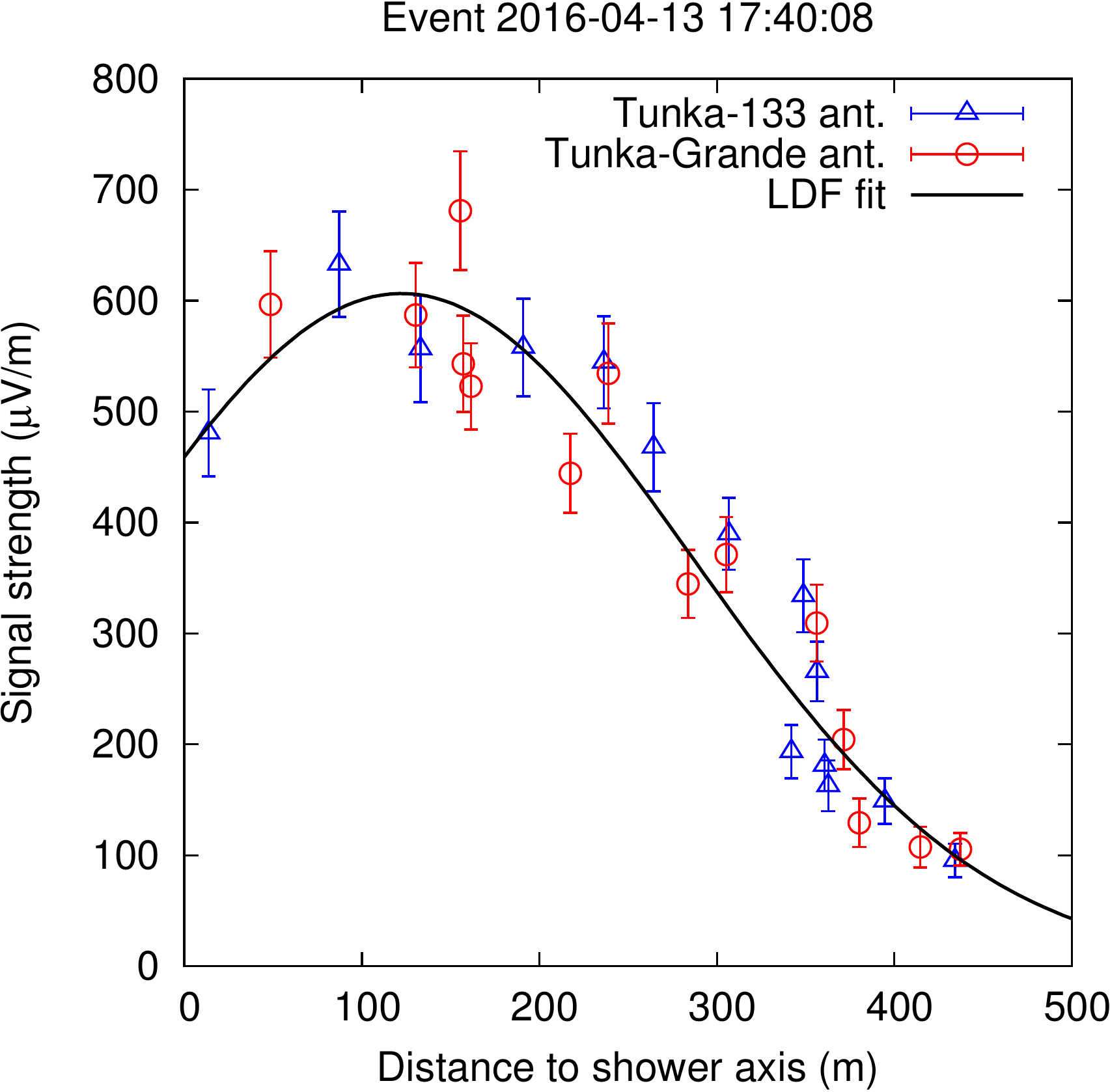} 
\caption{Lateral distribution of example event shown in FIG.~\ref{fig:event} with preliminary calibration of the antennas connected to Tunka-Grande. The curve indicates best LDF fit.}
\label{fig:ldf}
\end{figure}

During 2015-2016 the detection of air showers has been conducted by all TAIGA 
experiments except of the telescopes which are still under construction. 
The air-Cherenkov experiments Tunka-133 and Tunka-HiSCORE, whose operation is possible only in moonless nights, 
were operated during about 400 hours. The full duty-cycle experiments, detecting charged particles and 
radio emission from air-showers (Tunka-Grande and Tunka-Rex, respectively), operated during about 2000 hours. 
During summer (June-August), TAIGA was switched off because of the risk of damage by thunderstorms.
The analysis of the data measured jointly by the Tunka-Rex and Tunka-Grande experiments is still preliminary. 
We found about 2000 event candidates for energies above 100 PeV. An example of a reconstructed event is shown in FIG.~\ref{fig:event}. 
The lateral distribution function (LDF) of this example event is shown in FIG.~\ref{fig:ldf}. 
The methods of the combined Tunka-Rex and Tunka-Grande reconstruction are in progress.

\section{EFFECT OF ANTENNA ALIGNMENT}
Besides the main goals we studied the influence of the antenna alignment for cosmic-ray measurements. 
The radio signal from cosmic ray air showers is predominantly polarized along the geomagnetic Lorentz force, 
whose direction depends on the direction of the shower axis and the geomagnetic field 
(it corresponds to east-west direction if the shower axis is vertical)~\cite{KahnLerche1966}, 
thus, we had supposed that the efficiency of a radio detector should depend on the antenna alignment. 
This is the reason why the Tunka-Rex antennas are rotated by $45^\circ$ to the geomagnetic north-south axis, 
since we wanted to have more antennas with signal in both channels, tolerating to have less events 
with signal in at least one channel. 
To check the correctness of this assumption the vector product of the shower axis vector and the 
magnetic field vector $\bf{v} \times \bf{B}$ is used, where $\bf{v}$ is the shower axis vector and $\bf{B}$ is the magnetic field vector. 
Two configurations of antennas were considered: first, when antennas are aligned strictly along the north-south 
and east-west axes, which used in most experiments for detection of radio emission A($\bf{Ch_1}$, $\bf{Ch_2}$), and second, 
the alignment as in Tunka-Rex and LOFAR where antennas are rotated by $45^\circ$ with respect to the 
geomagnetic north-south axis A$^\prime$($\bf{Ch^\prime_1}$, $\bf{Ch^\prime_2}$). The next step is to evaluate the efficiency of each configuration 
by taking the projection of the vector product $\bf{v} \times \bf{B}$ on antenna arms for both alignments. Then, 
the difference between configurations was quantified by the difference between the highest amplitude 
in channels for each configuration (see Eq.~\ref{eq_1}):

\begin{eqnarray}
\Delta\bf{max}(\theta, \phi) = max[(\bf{E}, \bf{Ch_1})^2, (\bf{E}, \bf{Ch_2})^2] - {}\nonumber\\
- \bf{max}[(\bf{E}, \bf{Ch^\prime_1})^2, (\bf{E}, \bf{Ch^\prime_2})^2],
\label{eq_1}
\end{eqnarray}

where $\theta$ is zenith angle of the shower axis, $\phi$ – its azimuth, and $\Delta\bf{max}(\theta, \phi)$ 
is the difference between configurations A and A$^\prime$, $\bf{E}$ = $\bf{v} \times \bf{B}$ is E-field vector of the radio 
signal emitted by the shower.
Large values of $\Delta\bf{max}$, mean that the alignment has a large effect on the number of events detected 
in at least one channel or in both channels, respectively. The average dependence on the antenna 
alignment can be seen as function of the zenith angle $\theta$ when $\Delta\bf{max}$ is integrated over the azimuth angle $\phi$ (see Eq.~\ref{eq_2}):

\begin{equation}
I(\theta)= \int_{0}^{2\pi}\Delta\bf{max}(\theta, \phi)d\phi
\label{eq_2}
\end{equation}

\begin{figure}
\includegraphics[width=1.0\linewidth]{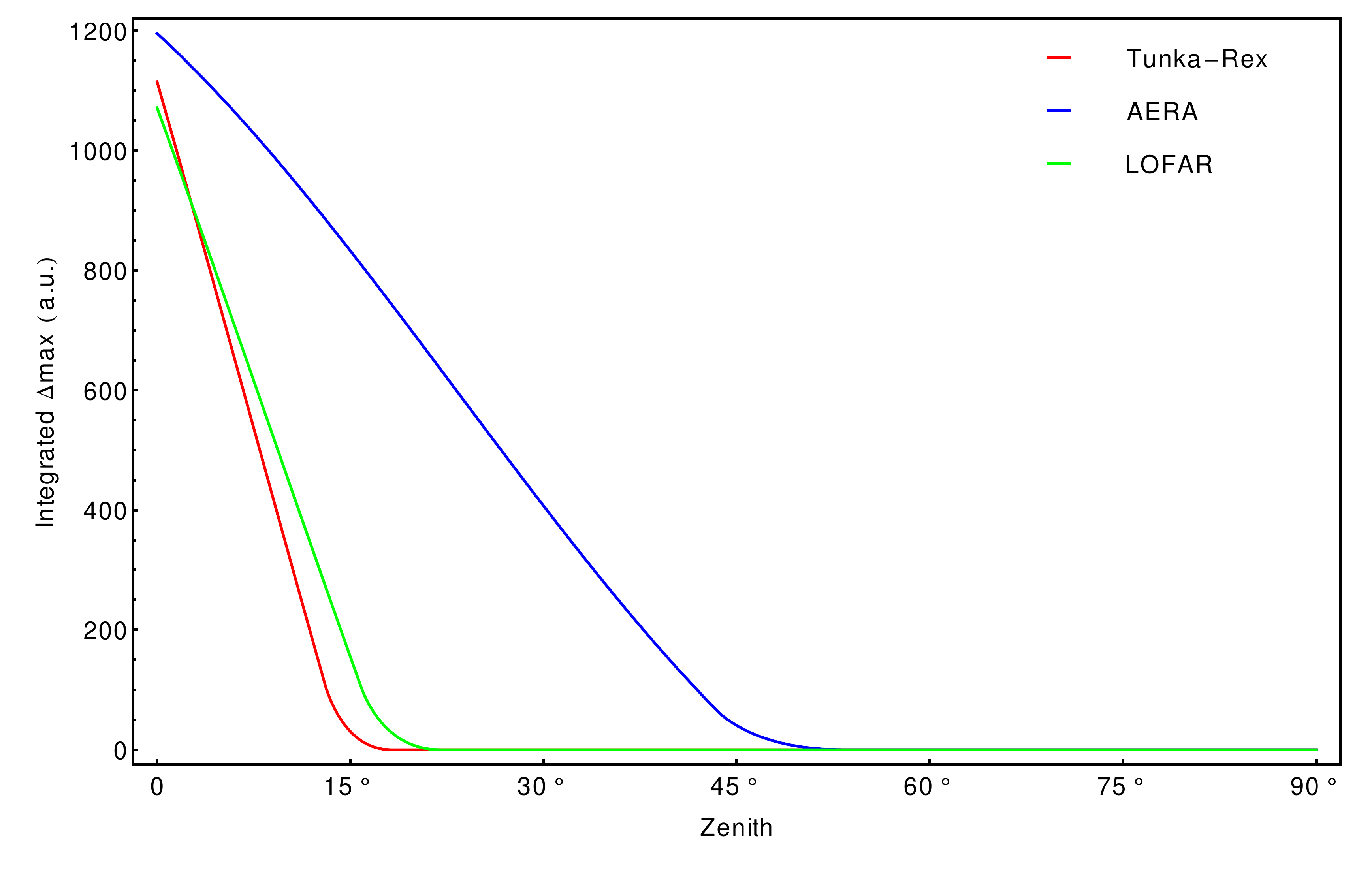} 
\caption{The effect of antenna alignment $I(\theta)$, see Eq.~\ref{eq_2}, depending on the shower zenith angle, for the geomagnetic fields at Tunka-Rex, AERA and LOFAR.}
\label{fig:idmax}
\end{figure}

FIG.~\ref{fig:idmax} shows the results of the calculation made for the geomagnetic field of various modern radio 
experiments: Tunka-Rex, AERA and LOFAR. From FIG.~\ref{fig:idmax} it can be concluded that the dependence 
on alignment vanishes on average when the zenith angle of an air shower $\theta$ is greater or equal to the 
geomagnetic zenith (i.e. inclination in the coordinate system of EAS) $\theta_{\bf B}$ 
(for Tunka-Rex $\theta_{\bf B}$ = $18.2^\circ$, 
for AERA $\theta_{\bf B}$ = $53.4^\circ$, and for LOFAR $\theta_{\bf B}$ = $22^\circ$). 
For an individual shower, the alignment can still be 
important, but taking into account that the shower directions are equally distributed over azimuth, 
the choice of antenna alignment becomes unimportant for zenith angles $\theta > \theta_{\bf B}$. This implies that for 
Tunka-Rex the antenna alignment is important only for vertical events (when the zenith angle of the 
shower is smaller than the geomagnetic zenith of $18.2^\circ$), and for more inclined events it does not matter. 
Only for the AERA experiment the choice of antenna alignment is important for a significant range of zenith 
angles, because the geomagnetic field at AERA is much more inclined.

The effect of antenna alignment was checked on the reconstructed Tunka-Rex data of 2012-2013. 
The total time of measurements was about 400 hours. The number of reconstructed events is 146 and 
there is only one event with zenith angle below the geomagnetic one. Thus, no significant effect 
is expected of the antenna alignment on the detection efficiency for Tunka-Rex. 
Furthermore, the effect of the antenna alignment on the efficiency of Tunka-Rex was studied by using 
CoREAS simulations~\cite{Huege:2013vt} in the presence of noise. About 300 events for proton and 300 events for iron 
as primary particle (where the energy and direction were taken from Tunka-133) were simulated. For 
these events experimentally measured noise was added. It was confirmed that neither the number of 
events nor the number of stations depends significantly on the azimuthal alignment of the antennas. 
Consequently, for Tunka-Rex either choice of antenna alignment is fine, as long as the exact alignment 
is known and can be taken into account during data analysis.

\section{CONCLUSION}
Tunka-Rex is the radio extension of the TAIGA experiment, a growing infrastructure for all components 
of air showers produced by high-energy cosmic rays as well as gamma-ray astronomy. 
Tunka-Rex provides competitive precision of the energy and the position of the shower maximum. 
Due to the absolute calibration of the radio antennas, the energy scales of KASCADE-Grande and Tunka-133 
could be compared and were found consistent within 10\%, which enables a better comparison of features 
observed in the energy spectrum.
After preliminary analysis of the experimental data, obtained by coincident measurements of Tunka-Rex 
and Tunka-Grande, we found about 2000 event candidates. The development of reconstruction methods is 
in progress and further Tunka-Rex analyses  will focus on a mass-composition study jointly with Tunka-Grande. 
Studies of the antenna alignment show that the efficiency of a radio detector does not dependent on 
the antenna orientation when the air shower zenith angle is larger than the geomagnetic zenith at the 
detector location. In particular for future arrays aiming at inclined air showers, this gives the 
freedom to choose the antenna orientation based on any other technical criteria. 

\bigskip % extra skip inserted
\begin{acknowledgments}
The construction of Tunka-Rex was funded by the German Helmholtz Association and the Russian Foundation 
for Basic Research (grant HRJRG-303). Moreover, this work was supported by the Helmholtz Alliance for 
Astroparticle Physics (HAP), by Deutsche Forschungsgemeinschaft (DFG) grant SCHR 1480/1-1, and by the 
Russian Federation Ministry of Education and Science (agreement 14.B25.31.0010). Also, this work 
was supported by the Russian Fund of Basic research grants 16-32-00329 and 16-02-00738.
\end{acknowledgments}

\bigskip % extra skip inserted
\bibliography{references.bib}

\begin{thebibliography}{20}
\expandafter\ifx\csname natexlab\endcsname\relax\def\natexlab#1{#1}\fi
\expandafter\ifx\csname bibnamefont\endcsname\relax
  \def\bibnamefont#1{#1}\fi
\expandafter\ifx\csname bibfnamefont\endcsname\relax
  \def\bibfnamefont#1{#1}\fi
\expandafter\ifx\csname citenamefont\endcsname\relax
  \def\citenamefont#1{#1}\fi
\expandafter\ifx\csname url\endcsname\relax
  \def\url#1{\texttt{#1}}\fi
\expandafter\ifx\csname urlprefix\endcsname\relax\def\urlprefix{URL }\fi
\providecommand{\bibinfo}[2]{#2}
\providecommand{\eprint}[2][]{\url{#2}}

\bibitem[{\citenamefont{{W. D.~Apel, et al.~(KASCADE-Grande
  Collaboration)}}(2012)}]{Apel2012183}
\bibinfo{author}{\bibnamefont{{W. D.~Apel, et al.~(KASCADE-Grande
  Collaboration)}}}, \bibinfo{journal}{Astroparticle Physics}
  \textbf{\bibinfo{volume}{36}}, \bibinfo{pages}{183} (\bibinfo{year}{2012}).

\bibitem[{\citenamefont{Sveshnikova et~al.}(2014)\citenamefont{Sveshnikova,
  Kuzmichev, Korosteleva, Prosin, and Ptuskin}}]{Sveshnikova:2015vga}
\bibinfo{author}{\bibfnamefont{L.}~\bibnamefont{Sveshnikova}},
  \bibinfo{author}{\bibfnamefont{L.}~\bibnamefont{Kuzmichev}},
  \bibinfo{author}{\bibfnamefont{E.}~\bibnamefont{Korosteleva}},
  \bibinfo{author}{\bibfnamefont{V.}~\bibnamefont{Prosin}}, \bibnamefont{and}
  \bibinfo{author}{\bibfnamefont{V.}~\bibnamefont{Ptuskin}},
  \bibinfo{journal}{Nucl.Phys.Proc.Suppl.} \textbf{\bibinfo{volume}{256-257}},
  \bibinfo{pages}{218} (\bibinfo{year}{2014}).

\bibitem[{\citenamefont{{F.G.~Schr\"oder}}(2017)}]{SchroederReview2016}
\bibinfo{author}{\bibnamefont{{F.G.~Schr\"oder}}}, \bibinfo{journal}{Prog.
  Part. Nucl. Phys.} \textbf{\bibinfo{volume}{93}}, \bibinfo{pages}{1}
  (\bibinfo{year}{2017}).

\bibitem[{\citenamefont{{N.~M. Budnev, et al.~- TAIGA
  Coll.}}(2014)}]{TAIGA_2014}
\bibinfo{author}{\bibnamefont{{N.~M. Budnev, et al.~- TAIGA Coll.}}},
  \bibinfo{journal}{JINST} \textbf{\bibinfo{volume}{9}},
  \bibinfo{pages}{C09021} (\bibinfo{year}{2014}).

\bibitem[{\citenamefont{Antokhonov et~al.}(2011)}]{Antokhonov:2011zza}
\bibinfo{author}{\bibfnamefont{B.~A.} \bibnamefont{Antokhonov}}
  \bibnamefont{et~al.}, \bibinfo{journal}{Nucl. Instrum. Meth.}
  \textbf{\bibinfo{volume}{A 628}}, \bibinfo{pages}{124}
  (\bibinfo{year}{2011}).

\bibitem[{\citenamefont{Budnev et~al.}(2015)\citenamefont{Budnev, Ivanova,
  Kalmykov, Kuzmichev, Sulakov, and Fomin}}]{Budnev:2015xca}
\bibinfo{author}{\bibfnamefont{N.~M.} \bibnamefont{Budnev}},
  \bibinfo{author}{\bibfnamefont{A.~L.} \bibnamefont{Ivanova}},
  \bibinfo{author}{\bibfnamefont{N.~N.} \bibnamefont{Kalmykov}},
  \bibinfo{author}{\bibfnamefont{L.~A.} \bibnamefont{Kuzmichev}},
  \bibinfo{author}{\bibfnamefont{V.~P.} \bibnamefont{Sulakov}},
  \bibnamefont{and} \bibinfo{author}{\bibfnamefont{{\relax Yu}.~A.}
  \bibnamefont{Fomin}}, \bibinfo{journal}{Moscow Univ. Phys. Bull.}
  \textbf{\bibinfo{volume}{70}}, \bibinfo{pages}{160} (\bibinfo{year}{2015}),
  \bibinfo{note}{[Vestn. Mosk. Univ.2015,no.2,80–84(2015)]}.

\bibitem[{\citenamefont{{P.A.~Bezyazeekov, et al.~- Tunka-Rex
  Coll.}}(2015)}]{TunkaRex_NIM_2015}
\bibinfo{author}{\bibnamefont{{P.A.~Bezyazeekov, et al.~- Tunka-Rex Coll.}}},
  \bibinfo{journal}{Nucl. Inst. Meth. A} \textbf{\bibinfo{volume}{802}},
  \bibinfo{pages}{89} (\bibinfo{year}{2015}).

\bibitem[{\citenamefont{{O.~Kr\"omer et al.~(LOPES
  Collaboration)}}(2009)}]{KroemerSALLAIcrc2009}
\bibinfo{author}{\bibnamefont{{O.~Kr\"omer et al.~(LOPES Collaboration)}}},
  \bibinfo{journal}{Proc. of the 31st ICRC, {\L}\'{o}d\'{z}, Poland}
  (\bibinfo{year}{2009}),
  \bibinfo{note}{{http://icrc2009.uni.lodz.pl/proc/html/}}.

\bibitem[{\citenamefont{Abreu et~al.}(2012)}]{Abreu:2012pi}
\bibinfo{author}{\bibfnamefont{P.}~\bibnamefont{Abreu}} \bibnamefont{et~al.}
  (\bibinfo{collaboration}{Pierre Auger}), \bibinfo{journal}{JINST}
  \textbf{\bibinfo{volume}{7}}, \bibinfo{pages}{P10011} (\bibinfo{year}{2012}),
  \eprint{1209.3840}.

\bibitem[{\citenamefont{{P.A.~Bezyazeekov, et al.~- Tunka-Rex
  Coll.}}(2016)}]{Bezyazeekov:2015ica}
\bibinfo{author}{\bibnamefont{{P.A.~Bezyazeekov, et al.~- Tunka-Rex Coll.}}},
  \bibinfo{journal}{JCAP} \textbf{\bibinfo{volume}{01}}, \bibinfo{pages}{052}
  (\bibinfo{year}{2016}).

\bibitem[{\citenamefont{{Kostunin} et~al.}(2016)}]{Kostunin:2015taa}
\bibinfo{author}{\bibfnamefont{D.}~\bibnamefont{{Kostunin}}}
  \bibnamefont{et~al.}, \bibinfo{journal}{Astropart. Phys.}
  \textbf{\bibinfo{volume}{74}}, \bibinfo{pages}{79} (\bibinfo{year}{2016}).

\bibitem[{\citenamefont{{R.~Hiller, et al.~- Tunka-Rex
  Coll.}}(2016)}]{HillerArena2016}
\bibinfo{author}{\bibnamefont{{R.~Hiller, et al.~- Tunka-Rex Coll.}}},
  \bibinfo{journal}{EPJ Web of Conferences} \textbf{\bibinfo{volume}{proc. of
  ARENA 2016, in press}} (\bibinfo{year}{2016}), \eprint{arxiv:1611.09614}.

\bibitem[{\citenamefont{{Pierre Auger Coll.}}(2016)}]{Aab:2015vta}
\bibinfo{author}{\bibnamefont{{Pierre Auger Coll.}}}, \bibinfo{journal}{Phys.
  Rev. D} \textbf{\bibinfo{volume}{93}}, \bibinfo{pages}{122005}
  (\bibinfo{year}{2016}).

\bibitem[{\citenamefont{{A.~Nelles et
  al.}}(2015)}]{NellesLOFAR_calibration2015}
\bibinfo{author}{\bibnamefont{{A.~Nelles et al.}}}, \bibinfo{journal}{JINST}
  \textbf{\bibinfo{volume}{10}}, \bibinfo{pages}{P11005}
  (\bibinfo{year}{2015}).

\bibitem[{\citenamefont{Buitink et~al.}(2014)}]{Buitink:2014eqa}
\bibinfo{author}{\bibfnamefont{S.}~\bibnamefont{Buitink}} \bibnamefont{et~al.},
  \bibinfo{journal}{Phys. Rev.} \textbf{\bibinfo{volume}{D 90}},
  \bibinfo{pages}{082003} (\bibinfo{year}{2014}).

\bibitem[{\citenamefont{et~al.
  (KASCADE-Grande~Collaboration)}(2010)}]{Apel:2010zz}
\bibinfo{author}{\bibfnamefont{W.~D.~A.} \bibnamefont{et~al.
  (KASCADE-Grande~Collaboration)}}, \bibinfo{journal}{Nucl. Instrum. Meth.}
  \textbf{\bibinfo{volume}{A 620}}, \bibinfo{pages}{202}
  (\bibinfo{year}{2010}).

\bibitem[{\citenamefont{{W.D.~Apel, et al.~- LOPES
  Coll.}}(2014)}]{2014ApelLOPES_MassComposition}
\bibinfo{author}{\bibnamefont{{W.D.~Apel, et al.~- LOPES Coll.}}},
  \bibinfo{journal}{Phys. Rev. D} \textbf{\bibinfo{volume}{90}},
  \bibinfo{pages}{062001} (\bibinfo{year}{2014}).

\bibitem[{\citenamefont{{W.D.~Apel, et al.~- LOPES and Tunka-Rex
  Colls.}}(2016)}]{TunkaRexScale2016}
\bibinfo{author}{\bibnamefont{{W.D.~Apel, et al.~- LOPES and Tunka-Rex
  Colls.}}}, \bibinfo{journal}{Physics Lett. B} \textbf{\bibinfo{volume}{763}},
  \bibinfo{pages}{179} (\bibinfo{year}{2016}).

\bibitem[{\citenamefont{{Kahn} and {Lerche}}(1966)}]{KahnLerche1966}
\bibinfo{author}{\bibfnamefont{F.~D.} \bibnamefont{{Kahn}}} \bibnamefont{and}
  \bibinfo{author}{\bibfnamefont{I.}~\bibnamefont{{Lerche}}}, in
  \emph{\bibinfo{booktitle}{Proceedings of the Royal Society of London. Series
  A, Mathematical and Phys. Sciences}} (\bibinfo{year}{1966}), vol.
  \bibinfo{volume}{289}, p. \bibinfo{pages}{206}.

\bibitem[{\citenamefont{Huege et~al.}(2013)\citenamefont{Huege, Ludwig, and
  James}}]{Huege:2013vt}
\bibinfo{author}{\bibfnamefont{T.}~\bibnamefont{Huege}},
  \bibinfo{author}{\bibfnamefont{M.}~\bibnamefont{Ludwig}}, \bibnamefont{and}
  \bibinfo{author}{\bibfnamefont{C.}~\bibnamefont{James}},
  \bibinfo{journal}{AIP Conf.Proc.} \textbf{\bibinfo{volume}{1535}},
  \bibinfo{pages}{128} (\bibinfo{year}{2013}), \eprint{1301.2132}.

\end{thebibliography}

\end{document}